# Deflection of High-Energy Cosmic Radiation Ions using a Bent Crystal Shield

M. B. H. Breese

*Department of Physics, National University of Singapore, Singapore 117542*

## ABSTRACT

A bent crystal shield which is capable of deflecting cosmic radiation ions of all atomic numbers away from spacecraft using the ion channeling effect is described. Its effect on the transmitted angular distribution of heavy, high-energy ions is characterized using Monte Carlo channeling simulations. Ions up to an energy limit set by the bent crystal curvature radius are deflected by the full curvature angle. A bent crystal shield with a 1×1 cm$^2$ surface area capable of deflecting such ions with energies up to 100 GeV/nucleon is fabricated in which the lattice planes across the entrance surface are uniformly aligned.

There is a low intensity, isotropic background of galactic cosmic radiation ions originating from supernovae[1,2] with the fluence peaking in the range of 100 to 1000 MeV/nucleon, with energies extending beyond 1 TeV/nucleon as shown in Fig. 1a. The elemental distribution decreases with atomic number with an additional peak at $^{56}$Fe [3]. Such heavy, high-energy (HZE) ions have a large linear energy transfer (LET) of hundreds of keV/μm. Their fluence is much lower than that of protons but HZE ions cause comparable amounts of damage to biological tissue owing to their higher LET resulting in damage to DNA, elevated cancer risks and genetic mutations. Ref. [4] reviews the effects of HZE ions on biological tissue as relevant to manned missions to Mars. Such extended durations of many months outside Earth's protective atmosphere and magnetic field mean that astronauts are exposed to the full cosmic and solar radiation backgrounds and suffer short and long-term exposure effects.

The problem is serious because HZE ions are impossible to fully shield against owing to their very high energies. A thick 'passive' radiation shield of 20 g/cm$^2$ stops ions up to energies of ~200 MeV/nucleon but results in the production of secondary radiation such as heavy secondary ions and nuclear recoils, high-energy photons, secondary electrons and neutrons. Even 2 g/cm$^2$ of shielding can result in the production of more secondary than primary ions, some of which have higher LETs. Secondary electrons form a low-LET background radiation which can also damage DNA and produce mutations [3]. The detailed effects of HZE ions on biological tissue and the adverse effects of shielding are under study [5,6]. Shields made of high Z elements tend to result in more secondary ion radiation from break-up of their large nuclei [5]. Shields made of low Z elements, such as liquid hydrogen or methane, are more effective in reducing the production of secondary radiation since they do not fragment into secondary particles [7], with proposals to shield spacecraft by enclosing them within liquid fuel tanks. Other forms of shielding spacecraft from HZE ions involve 'active' methods using electromagnetic deflection using high voltages or currents [8,9]. These approaches have difficulties sustaining the required voltages, high power and the unknown effects of subjecting astronauts to high magnetic fields over extended periods.

This paper describes how spacecraft may be shielded from ions of all atomic number originating from a narrow angular range using channeling in bent crystals, where the lattice field can deflect charged particles through small radii equivalent to a magnetic field strength of >1000 Teslas. Bent crystal channeling is normally used to deflect, extract and collimate high-energy charged particle beams in accelerators [10-15]. Protons with energies as low as 2 MeV [16] and as high as 900 to 980 GeV have been successfully deflected [17,18] with plans for use at 7 TeV [19,20]. Heavy ions such as 100 GeV/nucleon Au have also been deflected [21].

Bent crystal channelling occurs for those ions incident on the surface within the planar channeling critical angle, $\theta_c$:

$$\theta_c = \sqrt{\frac{4Z_1Z_2e^2 Nd_p Ca_{TF}}{pv}} \qquad (1)$$

where $Z_1$ and $Z_2$ are the atomic numbers of the incident and lattice nuclei, $N$ is the atomic density, $d_p$ in the planar spacing, $C \cong \sqrt{3}$, $a_{TF}$ is the Thomas-Fermi screening distance and $p,v$ are the ion momentum and velocity. If the curvature radius, $R$, of a bent lattice exceeds a critical value, $R_c$:

$$R_c = \frac{pv}{\pi Z_1 Z_2 e^2 Nd_p} \qquad (2)$$

then many planar channeled ions of a given energy are steered through the full curvature [10]. As a consequence of Lindhard's reversibility rule [22] a large angular distribution of ions transmitted through a bent crystal produces no net deflection [23]. Deflecting high-energy ions away using a bent crystal shield thus only applies to ions originating from a small angular range, such as a single supernova which occur in Milky Way-type galaxies about once every 30 years. All channeling simulations in this paper use a parallel beam of HZE ions.

Detailed simulations of the bending effect of curved silicon (110) lattice planes were performed for different ions with energies up to 500 GeV/nucleon using the Monte Carlo code FLUX [24] which uses a binary collision model in conjunction with the Ziegler-Biersack-Littmark potential [25]. Since the channeling behaviour of all fully-stripped ions at the same energy/nucleon is similar only the results for $^{56}$Fe are discussed.

Fig. 1b shows angular distributions of planar and randomly aligned 200 MeV/nucleon Fe ions transmitted through a lattice bent by 5 mrad. Many planar aligned ions are deflected (dashed lines in inset) to an exit angle of –5 mrad and emerge within a narrow angular range of $\pm\theta_c$. The deflected beam fraction of 73% is determined by the beam fraction of 25 to 30% which does not undergo planar channeling at the entrance surface [26]. Nonchanneled and randomly aligned ions undergo multiple scattering about their initial direction with no net deflection (solid lines in inset). Fig. 1c shows angular distributions for a range of planar aligned Fe ion energies through the same curved lattice. At high ion energies $R < R_c$ so the curvature radius is too small for the lattice field to maintain the ions within the planar channels so they are transmitted in a random manner. At intermediate energies $R > R_c$ so most ions are deflected. At lower energies a smaller fraction is deflected through the full curvature owing to increased planar dechanneling. Such angular distributions were simulated for a wide range of ion energies in layers bent through 5 mrad with different bending radii and thicknesses. The bending efficiencies shown in Fig. 1d for deflection in layers bent by different curvature radii decreases to zero at a certain energy but high efficiency can be attained with a larger $R_c$. Fig. 1d also shows the value of $\theta_c$ to indicate the angular range over which high bending efficiency is achieved at different energies. It decreases from 250 µrad (~1 minute of arc) at 200 MeV/nucleon to 5 µrad (~1 second of arc) at 1 TeV/nucleon.

The manner in which bent crystal shields can deflect cosmic HZE ions away from spacecraft is demonstrated in Fig. 2. This shows the combined deflecting effect of two bent crystal shields of differing curvature radii on equal intensities of 200 MeV/nucleon and 2 GeV/nucleon Fe ions approaching from the left. The thin first shield encountered has $R = 10$ mm and this deflects most of the lower energy ions whereas the higher energy ions are transmitted with little scattering. The thicker second shield has $R = 100$ mm and this deflects most of the higher energy ions resulting in the spacecraft being subjected to a lower fluence over a wide spectrum of ion energies. The effect of these combined shields on the energy spectrum of a parallel beam of Fe ions is shown in Fig. 1a though it applies to ions of all atomic numbers, resulting in a reduction of about 70% in the ion fluence irradiating the spacecraft.

This approach requires fabrication of suitable bent crystal shields with large areas and this is now considered. A bent crystal layer was previously produced over a six inch wafer surface by growing a graded SiGe layer on a [001] silicon substrate [16], and deflection of 2 to 5 MeV protons through a lattice with R = 50 µm was observed. Thicker graded $Si_{1-x}Ge_x$ layers were recently grown with R = 10 mm [27] which are capable of deflecting ions with energies up to ~1 GeV/nucleon. However, this approach is not suitable for bending higher ion energies owing to difficulties in accurately grading low Ge compositions and growth of very thick layers.

The usual approach to deflecting GeV ions using bent crystal channeling is to pass them through a planar aligned silicon wafer with a thickness of 0.5 to 1.0 mm, length of 3 mm to 20 cm, and bent through angles of 100 µrad to 10 mrad. A similar approach was used here to construct large-area bent crystal shields by bonding twenty [001] silicon wafers together with slow-setting epoxy, each 0.5 mm thick, 1 cm wide and 6 cm long. The wafers were bent through almost the same curvature radius and angle between wire-cut, polished stainless steel surfaces, one convex with a radius $R = 1.00$ m and the other concave with $R = 1.01$ m as shown in Fig. 3a. Once dry the bent crystals were cut into arc lengths of 5 mm and the entrance and exit faces polished to give surfaces across which the lattice planes have the same nominal alignment, as shown in Fig. 3b. This resultant bent crystal shields have surface areas of $1\times1 cm^2$ and $R = 1.00$ to 1.01 m, capable of deflecting HZE ions with energies up to ~100 GeV/nucleon by 5 mrad from Fig. 1d.

The curvature angle of the bent crystal shield was measured by aligning the vertically-running (110) planes perpendicular to the surface-normal of the entrance face with a 2 MeV proton beam ($\theta_c = 0.17°$) focused to a 2 μm spot in a nuclear microprobe [28,29]. Fig. 4a shows the measured planar channeling dip. The shield was then rotated through 180° and the angular location of the exit face relative to the entrance face was measured as 179.73±0.03°, giving a wafer bend angle of 4.7±0.4 mrad compared with a nominal angle of 5 mrad.

The shield was then tilted 0.10° away from alignment of the entrance face so the rapidly changing portion of the channeling dip was aligned with the incident beam, making subsequent measurements more sensitive to any variations in tilt angle of the surface planes. The focused proton beam was scanned horizontally across the 1 cm shield width and the backscattered yield at each lateral position was measured. Variations in the alignment of the (110) entrance planes across the surface are manifested as a change in the yield with lateral position [30,31]. Fig. 4b shows a typical result measured across the entrance surface, plotted as percentage variations of the yield where 100% corresponds to random alignment. The measured yield across the entire area is flat to within ±1% except for zones ~20 μm wide between adjacent wafers due to the epoxy layer and possible anticlastic bending at the wafer edges [11]. This demonstrates that a uniformly aligned entrance surface can be produced to within a measurement sensitivity of ~0.2 mrad, across 96% of the 1 cm shield width.

This work describes a new way of shielding spacecraft from HZE ions using bent crystal shields. Shield areas of $1\times1$ cm$^2$ have been fabricated and larger areas can be constructed using an array of such shields. The range of ion energies over which a shield is effective depends only on its curvature radius with no obvious upper limit to the energies which may be deflected. While only a parallel beam of ions incident on the shield and deflected through bent crystal channeling is considered here for simplicity, recent work has shown how a beam with an angular range many times the channeling critical angle can be deflected through a much larger angle by a related process called multiple volume reflection [32].

Lightweight shields may be constructed with areal densities of 10 mg/cm$^2$ and 1 g/cm$^2$ for deflecting ion energies up to 200 MeV/nucleon and 100 GeV/nucleon respectively, in comparison with passive shielding which require layers $10^3$ to $10^4$ times thicker to stop these same ions. Because the bent crystal shields are thin and located away from the spacecraft then any production and effects of secondary radiation will be minimal. While only the use of such shields for deflecting cosmic radiation has been described here, other possible uses may include large-area beam collimators and extractors, fast beam dumps and lightweight, portable shields.

**Figure Captions**

**Figure 1**

(a). Cosmic background ion radiation fluences for H, He, C, Fe, reproduced from Ref. [1]. The dotted line shows the calculated Fe fluence resulting from the two bent crystal shields shown in Fig. 2, assuming radiation is emitted from a single direction. (b) Planar and random aligned angular distributions of 200 MeV/nucleon Fe ions transmitted through a 50 μm thick layer with $R$ = 10 mm. The inset shows the bent crystal channeling geometry. (c) Angular distributions of planar aligned Fe ions of different energies through the same layer as (b). (d) Deflection efficiency (beam fraction which is deflected through the full bend angle and emerges within $\pm\theta_c$) as a function of Fe ion energy for four different bent layer curvature radii. The value of $\theta_c$ is shown by the dashed line and the right-hand scale.

**Figure 2**

FLUX simulation for two shields with $R$ = 10 mm and $R$ = 100 mm deflecting Fe ions approaching from the left, with energies of 200 MeV/nucleon (red) and 2 GeV/nucleon (black). The shields are located 5km and 10 km to the left of the spacecraft, with surface areas of 10×10 m$^2$, slightly larger than the cross-section of the spacecraft. The shields are not attached to the spacecraft and can be aligned independently.

**Figure 3**

(a). Fabrication of many bent crystal shields with surface areas of 1×1 cm$^2$ by bending 20 silicon wafers between concave and convex surfaces. The wafers were cut and polished along radii of the lattice curvature (dashed lines) with arc lengths of 5 mm. (b) Plan view of a shield entrance face with the (110) surface planes having the same alignment across full width.

**Figure 4**

(a). Planar channeling dip of Rutherford backscattered protons recorded from the vertically-running (110) planes of a bent crystal shield. (b). 750 μm linescan of the measured backscattered proton intensity across the shield entrance face with the shield tilted 0.10º away from alignment, to the arrowed location in (a). The error bars are based on the counting statistics at each pixel

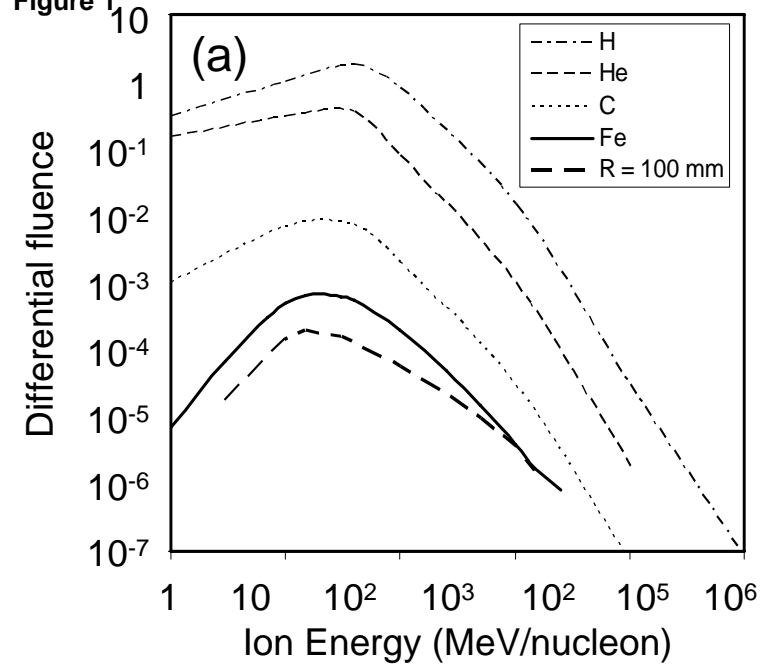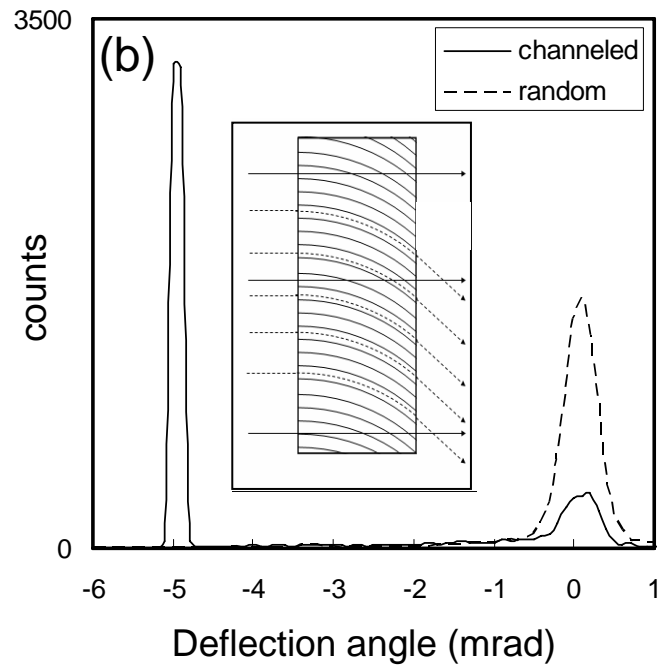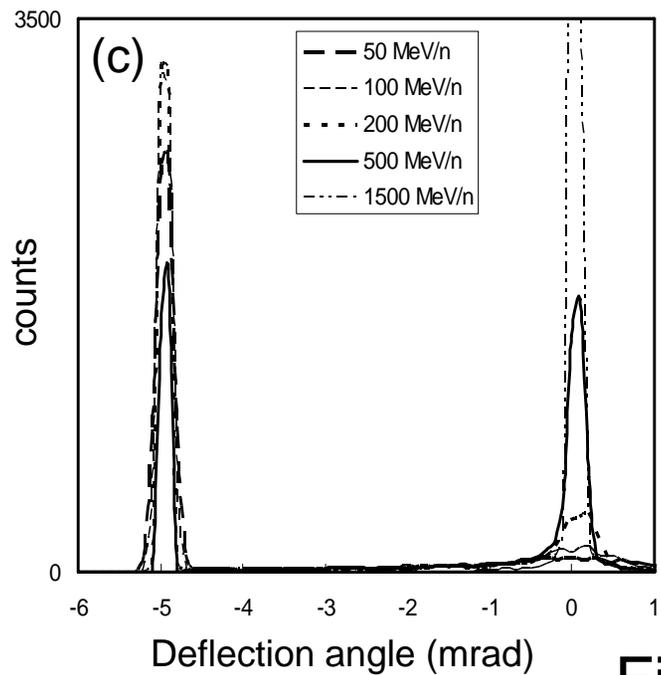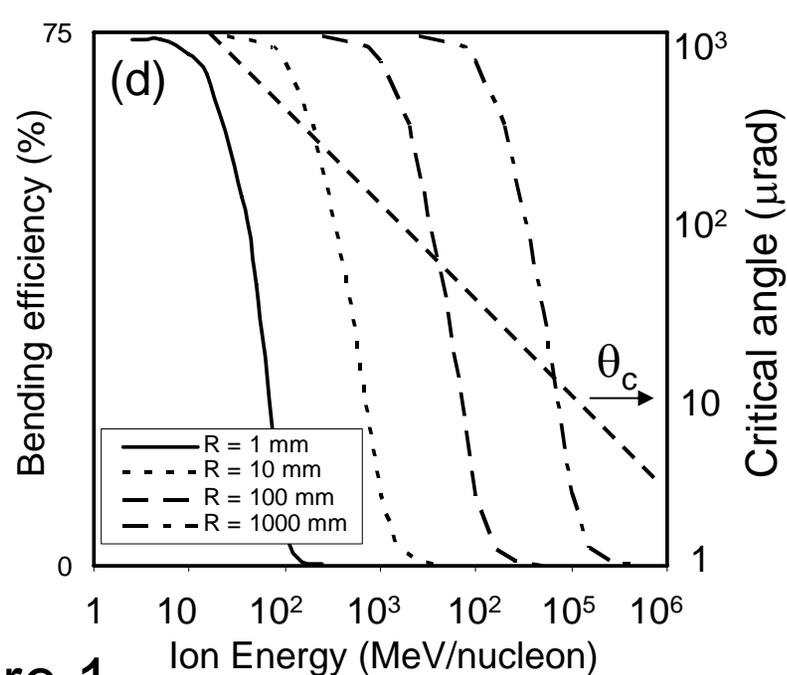

Figure 1

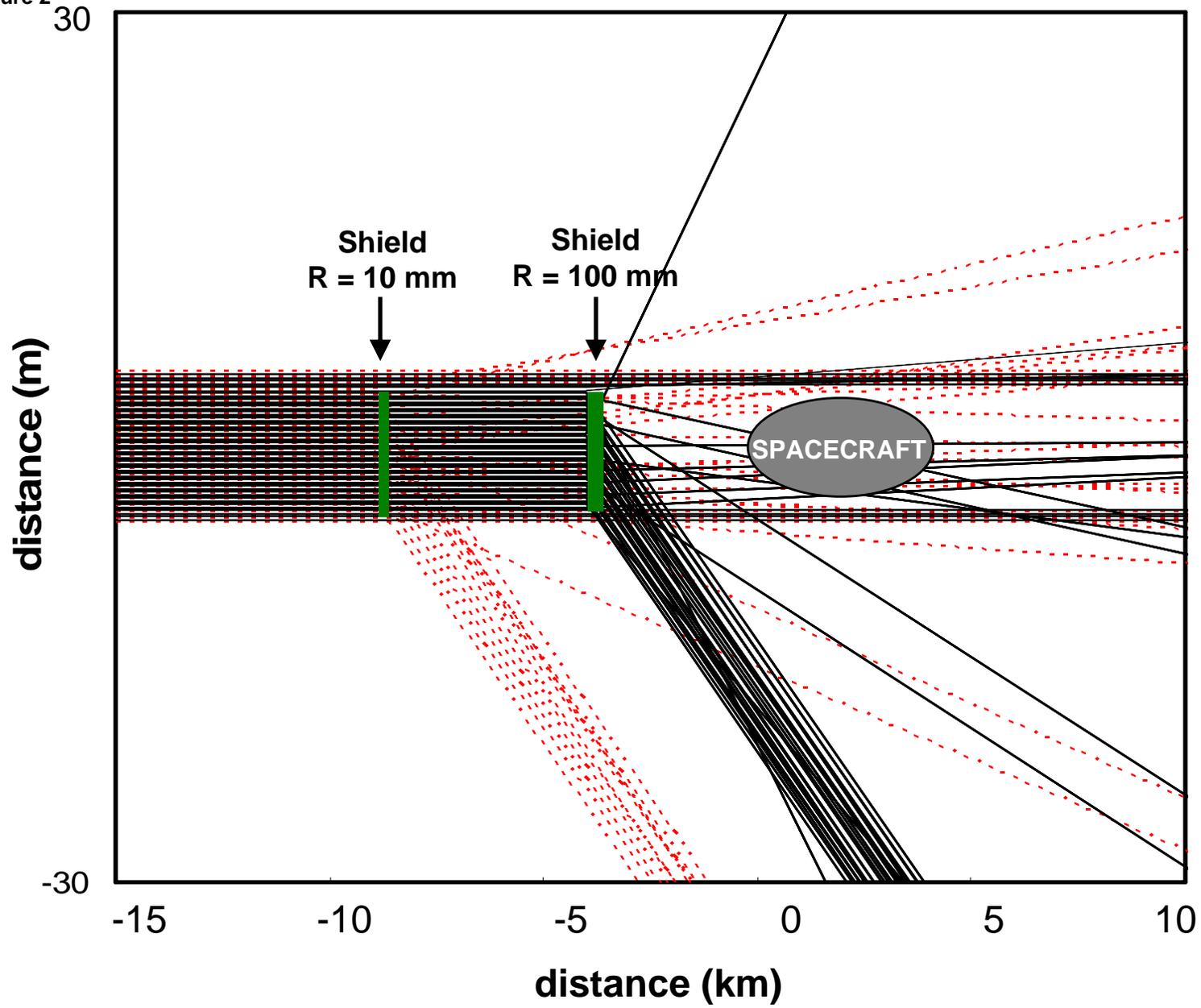

Figure 2

**Figure 3**

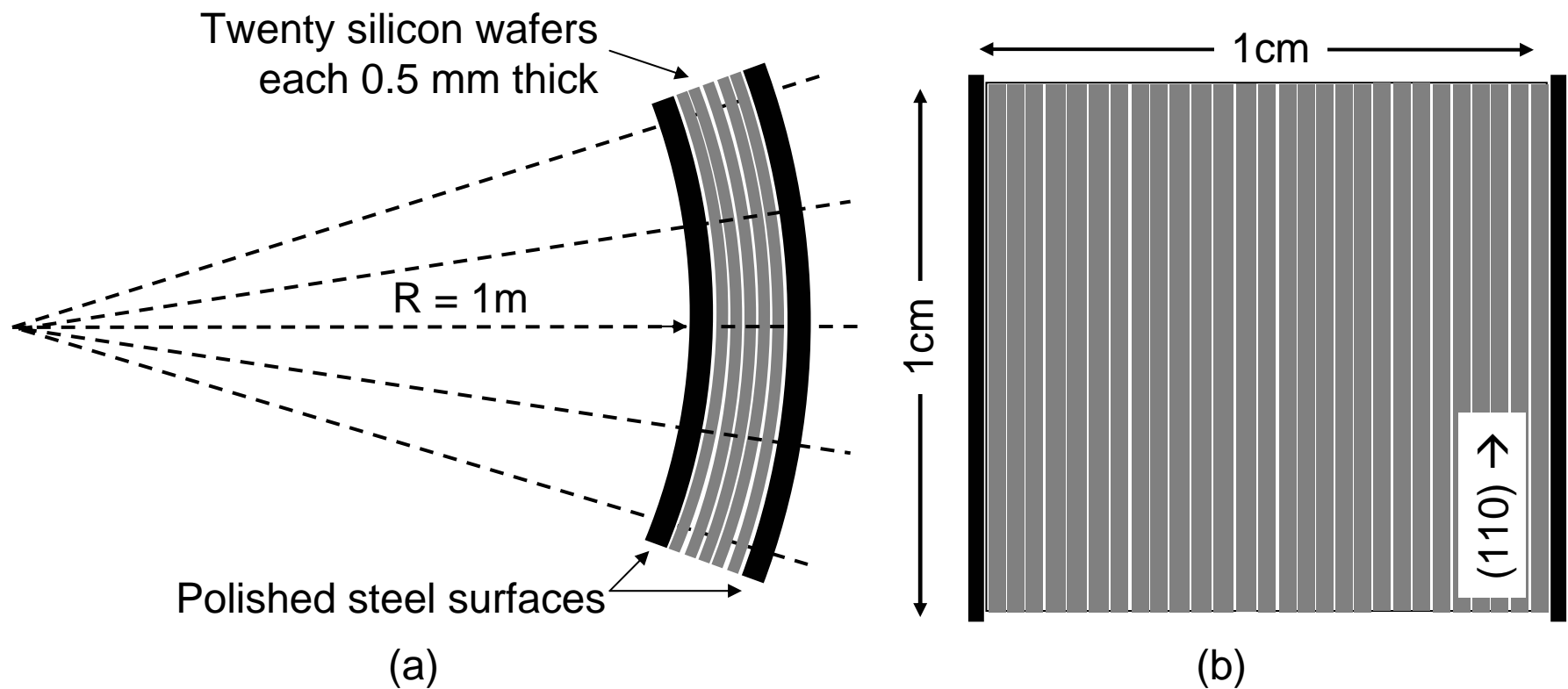

Figure 3

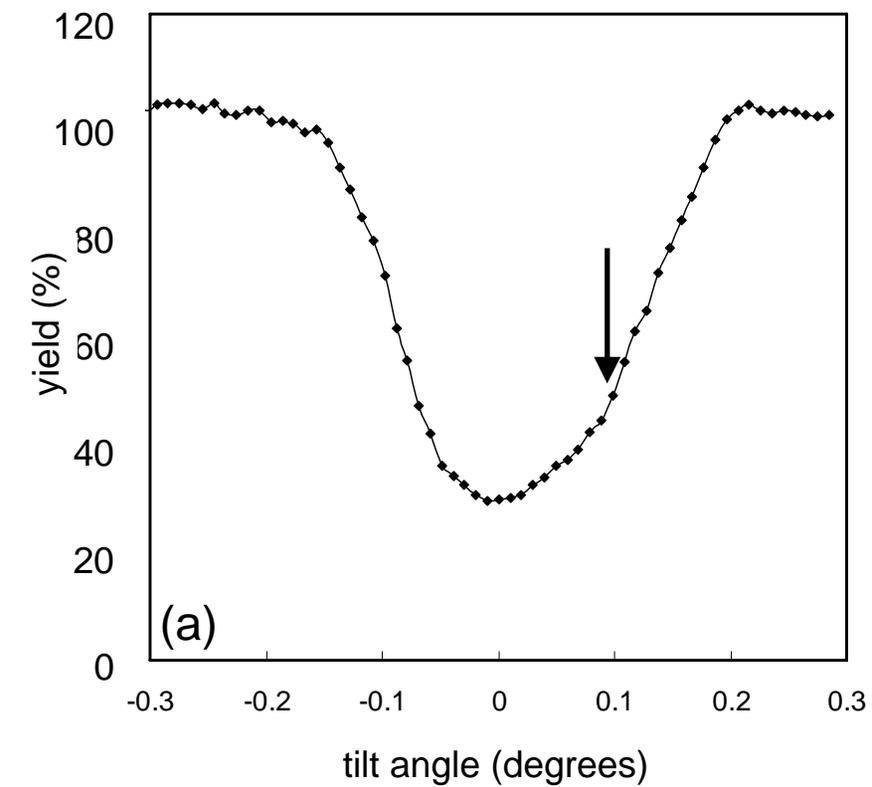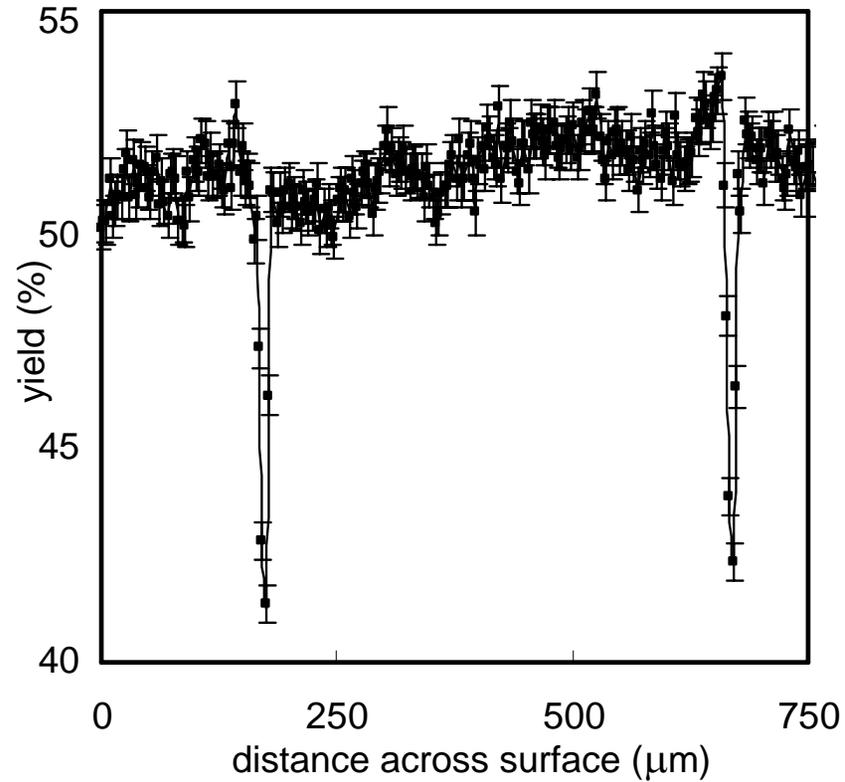

Figure 4